\documentclass[conference]{IEEEtran}

\usepackage{amsmath}
\usepackage{amsfonts}
\usepackage{amssymb}
\usepackage{multicol}
\usepackage[latin1]{inputenc} 

\ifCLASSINFOpdf
   \usepackage[pdftex]{graphicx}
   \DeclareGraphicsExtensions{.pdf,.jpeg,.png}
\else
   \usepackage[dvips]{graphicx}
   \DeclareGraphicsExtensions{.eps}
\fi

\usepackage[tight,footnotesize]{subfigure}

\begin{document}
%
\title{The Deletable Bloom filter \\ - A new member of the Bloom family -}

\author{\IEEEauthorblockN{Christian Esteve Rothenberg, Carlos A. B. Macapuna, 
F\'{a}bio L. Verdi$^\dag$ and Maur\'{i}cio  F. Magalh\~{a}es}
\IEEEauthorblockA{University of Campinas  (Unicamp), $^\dag$Federal University of S\~{a}o Carlos (UFSCar) \\
  \{chesteve, macapuna, mauricio\}@dca.fee.unicamp.br, $^\dag$verdi@ufscar.br}
}

\maketitle

\begin{abstract}
We introduce the Deletable Bloom filter (DlBF) as a new spin on the popular  data structure based on compactly encoding the information of where collisions happen when inserting elements. The DlBF design enables false-negative-free deletions at a fraction of the cost in memory consumption, which turns to be appealing for certain probabilistic filter applications. 
\end{abstract}


%
\IEEEpeerreviewmaketitle

\section{Introduction}
The Bloom filter (BF)~\cite{362692} is a popular data structure capable 
of answering questions of the form ``is element $x$ in set $S$?'', with some tunable 
probability of returning false positives, i.e., claiming that $x$ belongs to $S$ even when this is not true.
Due to its simplicity and wide applicability, BFs have become very interesting objects of study and a daily aid of system implementations.
The 40-year-old hash-based data structure 
is beloved by theoreticians due to the mathematics that underpin the randomized flipping of 0s into 1s, and is beloved by practitioners as a powerful ally when aggregating data sets. 
BFs turn resource-intense (memory, computation) operations into simple, resource-friendly set membership problems.
The Bloom domain spans from 
hardware implementations, 
 all the road up the system stack to the software application domain, where it first saw the light
to perform space- and time-efficient dictionary look-ups.
 
The design of BFs is fundamentally about tradeoffs, i.e., striking the right balance between memory, computation and (false positive) performance. 
 Several variations have been proposed to 
modify the behavior of the standard Bloom filter (SBF) beyond its natural limits, 
for instance, sacrifying its zero false negative characteristic in favor of less false positives, e.g.,~\cite{1368454}. 
Due to its broad scope of applications, such 
metamorphoses are commonly needed to meet application-specific requirements or render additional features like frequency queries, deletions, coding values, security, and so on.

In this letter, we give birth to a new BF spin-off: The Deletable Bloom filter (DlBF). The DlBF inherits the plainness of its progenitor and introduces only a simple yet powerful idea, namely keeping record of the bit regions where collisions happen. The proposed design tradeoff turns out to be useful for applications with the following requirements:

\begin{description}
\item[R1:] Probabilistic guarantees of element deletability.
\item[R2:] No false negatives upon element deletion.
\item[R3:] Fixed memory allocation. 
\item[R4:] Low impact on the false positive rate ($fpr$) i.e. comparable to a SBF of the same bit size $m$. 
\end{description}


Like other Bloom scions in the past, our needs for another Bloom variant come from a specific networking application 
(see in-packet BF examples in Sec.~\ref{sec:applications}).  
However, the DlBF is well suited for other use cases where re-constructing the filter 
 upon set membership changes is either unfeasible or too costly.
For standalone applications, removal of element fingerprints is commonly desirable for functionality or optimization purposes.
For distributed applications, a deletable filter can be 
 thinned out as queried elements are processed in order to 
(i) avoid repeated matches upfront, 
(ii) reduce false positives, 
and/or (iii) enable fresh bit space for new additions.

\section{Related work}
\label{sec:related}

The first Bloom descendant with genes for deletability is the Counting Bloom filter (CBF)~\cite{343572}, which basically extends the 1-bit cells to c-bit counters.
 Unfortunately, this c-fold reduction of practical bit space, typically 3 or 4 bits to avoid counter overflows, is a price too high in memory consumption (e.g., on-chip memory).  
Bloom 
 relatives that improve this waste of space 
  include the Spectral Bloom filter
~\cite{872787}, and ``an optimal Bloom filter replacement''
~\cite{1070548}. While proven by theory to be more space-efficient, both alternatives come with a non-negligible complexity overhead, missing thereby the implicit requirement of \textit{simplicity}, a critical factor for actual implementations. 
The d-left CBF (dlCBF)~\cite{Bonomi06animproved} is probably the best alternative construction for a CBF. Based on d-left hashing and element fingerprints, the dlCBF is simple, 
  and given a target $fpr$, it requires about half the bit-space $m$ of a 4-bit CBF. However, aiming at a $fpr$ comparable to a SBF (R4), we can not afford 
  around $2m$ for construction. 

Closest to our design, is the Bloom filter with variable-length signatures (VLF) by Lu et al.~\cite{Lu05bloomfilters}, which presents an elegant solution to the deletion problem by resetting only a fraction of the $k$ bits. 
 Unfortunately, the main caveat of the VLF is that it is prone to false negatives, missing thereby R2. 
To the best of our knowledge, there is no Bloom filter variation which simultaneously satisfies all requirements R1-R4. 



\section{Design}
\label{sec:design}

\begin{figure*}[t] 
       \begin{minipage}[b]{0.6\linewidth}
           \includegraphics[width=\linewidth]{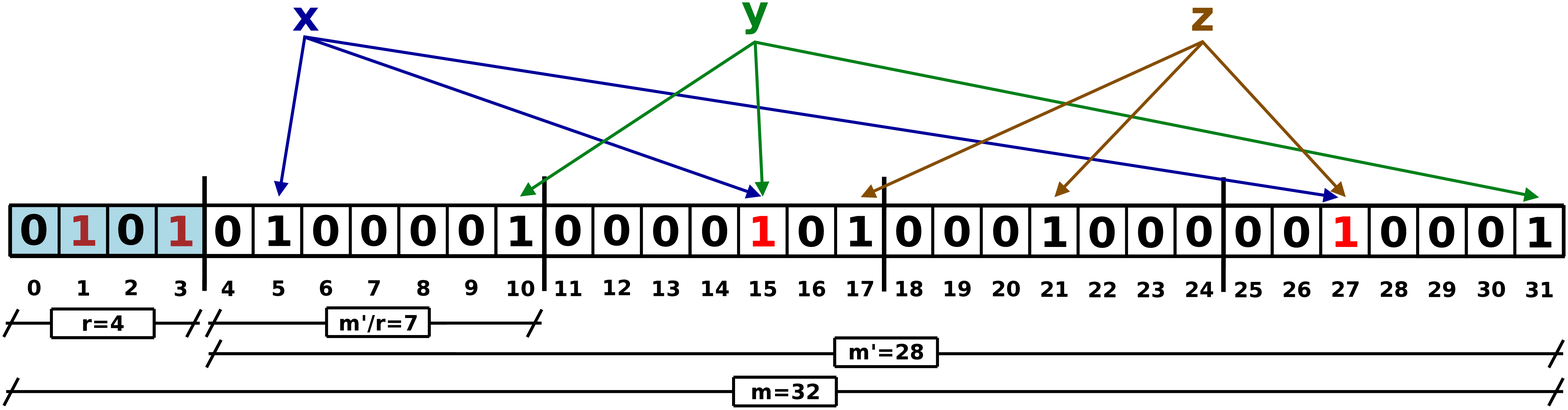}\\
           \caption{Example of a DlBF with $m=32$, $k=3$ and $r=4$, representing $S=\{x, y, z\}$. 
 The 1s in the first $r$ bits indicate collisions in the corresponding regions and bits therein cannot be deleted. All elements are deletable as each has at least one bit in a collision-free zone.}
           \label{fig:DlBF}
       \end{minipage}\hfill
       \begin{minipage}[b]{0.38\linewidth}
           \includegraphics[width=\linewidth]{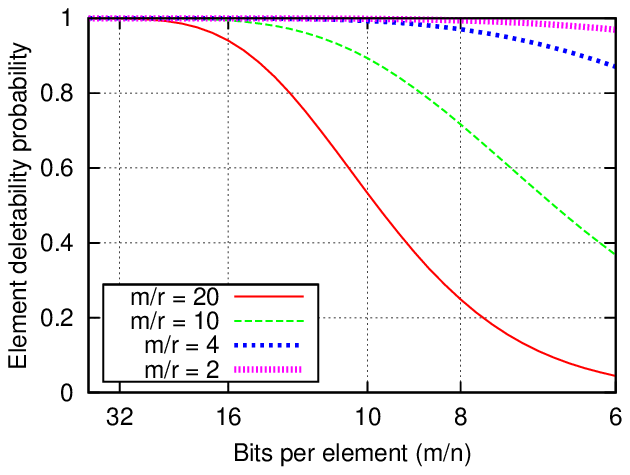}\\
           \caption{Deletability estimate as function of the filter density $m/n$ for different collision bitmap sizes $r$.}
           \label{fig:del-r}
       \end{minipage}\hfill
   \end{figure*}

The DlBF is built on the simple idea of tracking where bit collisions occur when inserting elements and exploits that bits set by only one element can be safely deleted. The proposed amendment consists of compactly encoding the regions of deletable bits using a fraction of the filter memory. 
An element can be effectively removed if at least one of its bits is reset, i.e., located in a collision-free-region.
 We divide a bit array of size $m$  into $r$ regions of $\lceil m'/r \rceil$ bits each, where $m'$ is the original $m$ minus the bits required to code the information of the collisions. A straightforward approach to compactly represent this information is a bitmap of size $r$ to code with $0$ a collision-free region 
and with $1$ otherwise (see Fig.~\ref{fig:DlBF}). 
 Element insertion and lookup are the same as in a traditional BF. In addition to adding and maintaining a collision bitmap of size $r$, the DlBF adds an element removal primitive:
\begin{itemize}
\item \textit{Insert(x)} maps an element $x$ to $k$ bit positions determined by a set of independent hashes. 
 If one bit cell happens to be already set (collision), the corresponding region is marked in the $r$ bitmap as non-deletable.
\item  \textit{Query(x)}  returns \textit{true} if the $k$ bit positions are set to 1. 
\item  \textit{Remove(x)} clears only those bit positions among $k$ which are located in collision-free zones. 
\end{itemize}
False-negatives are avoided at the cost of some elements not being deletable and accounting now as false positives, which are acceptable by the Bloom filter principle. 
 Orphaned (non-removable) bits contribute to a larger fill factor, which, in turn, deviates the observed $fpr$ from the expected value if all parameters were optimized. 
Consequently, one limitation of the DlBF appears in dynamic applications with frequent deletions and insertions where orphaned bits may fill the filter until collisions have happened in every region, hampering future deletions and increasing the residual $fpr$. 


Since element removal is only probabilistic, a key design issue is choosing the value $r$ and quantifying its impact on (i) the capacity to remove elements, and (ii) the false positive behavior (before and after elements are removed).
First, we provide the mathematical model for the element deletability probability and then we estimate the false positive penalties.


\subsection{Element deletability probability}

Consider a bit array of size $m'=m-r$ with $\lceil m'/r \rceil$ bit cells per region.   
The probability that a given cell has at least one collision is $p_c = 1-p_0-p_1$, where
$p_0$ denotes the probability that a given cell is set to 0 and $p_1$ is the
probability that a given cell is set to 1 only once after inserting $n$ elements: 
\begin{footnotesize}
\begin{equation*}
  p_{0} = (1-1/m')^{kn} \textrm{and }  p_{1} = (kn)(1/m')(1-1/m')^{kn - 1}
\label{eq:p}
\end{equation*}
\end{footnotesize}
Then, the probability that a $m '/ r$ bit region is collision-free is given by $(1-p_c)^{m '/ r}$. 
Finally, for $r \geq k$ and $m>>k$, the probability of an element being deletable (i.e., with one of its $k$ bits in a collision-free region) can be approximated to:
\begin{footnotesize}
\begin{equation}
  p_{d} = (1-(1-p_c)^{m '/ r})^k
\label{eq:pd}
\end{equation}
\end{footnotesize}
Figure~\ref{fig:del-r} plots $p_d$ against the filter density $m/n$ for different memory to regions ratios $m/r$, confirming the intuition that increasing $r$ results in a larger portion of deletable elements. As more elements are inserted (lower $m/n$), the number of collisions increase and consequently the deletion capabilities are reduced. 
Hence,  the parameter $r$ can be chosen by defining a target element deletion probability $p_d$ and estimating the upper bound of the set cardinality $n$. 
For instance, allocating only 5 \% of the available bits ($m/r=20$) to code the collision bitmap, we can expect to remove around 90 \% of the elements when the bits per element ratio $m/n$ is around 16. 
%
%
%

\subsection{False positive probability}

The false positive impact of consuming $r$ bits from $m$ 
can be estimated by updating $m$ in the well-known false positive probability of a BF: 


%
%
%
%
%

\begin{footnotesize}
\begin{equation}
  {p_r}^k = \left[1 - \left(1 - \frac{1}{m-r}\right)^{k*n}\right]^k \label{eq:Pdlbf}
\end{equation}
\end{footnotesize}
Obviously, the false positive degradation is driven by the ratio $m/ r$.  
With  $r$ being only a fraction of $m$, the false positive increase is controllable and arguably comparable to a SBF, satisfying thus $R4$ ($fpr_{m'=m-r} \approx fpr_{m}$).



\section{Practical evaluation}
\label{sec:evaluation}

We now evaluate via simulation the practical performance of the DlBF in terms of \textit{deletability} and observed $fpr$. 
We answer the questions of (1) how many elements can be safely removed in practice, and (2) how many false positives are observed before and after elements are removed. 

\begin{figure}[t]
\centering
\includegraphics[width=0.4\textwidth]{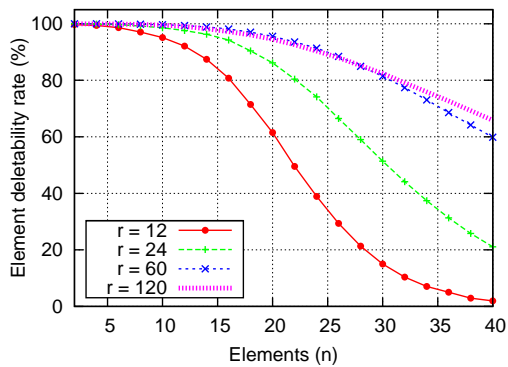} 
\caption{Experimental deletability rate (mean values) of a 240-bit DlBF with $k=5$ for different number of regions $r$.}
\label{fig:bfd-exp}
\end{figure}

\begin{figure}[t]
\centering
\includegraphics[width=0.4\textwidth]{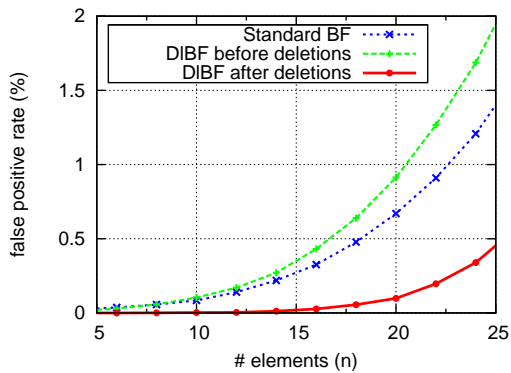} 
\caption{Observed average $fpr$ of a 240-bit DlBF with $r=24$ and $k=5$, before and after removing elements, compared to a 240-bit SBF.}
\label{fig:fpr-exp}
\end{figure}

Due to space limitations, we present only the experimental results for the case where $m = 240$ and $k=5$, which corresponds to the configuration of the in-packet BF application~\cite{lipsin} that motivated the DlBF design (see Sec.~\ref{sec:applications}). On every trial (2000 per parameter set), we insert $n$ elements randomly chosen from the American dictionary ($\approx$ 145K entries). We then (i) quantify how many inserted elements can be deleted (Fig.~\ref{fig:bfd-exp}), and (ii) count for false positives (Fig.~\ref{fig:fpr-exp}) by testing 500 randomly chosen elements (before and after deletions).


The observed deletability rate behaves as predicted by theory, but with relatively lower values (noteworthy as $r$ tends to $m/2$ and for high $m/n$ ratios). This can be explained by the assumption in Eq.~\ref{eq:pd} of perfectly random hash functions, an issue which can be more significant in small BFs~\cite{bose2008false}. 
Taking as an example the case where 10\% of the memory is used to code the bitmap ($r=24$), under a reasonably utilization ($n=22$), on average, 80\% of the elements could be removed (compared to 90\% predicted by theory)  by resetting around 40\% of the bits (not shown in Fig.~\ref{fig:bfd-exp}). Interestingly, doubling $r$ from 60 to 120 only improves the number of deleted bits but not the actual element deletability. 
As expected, the price in $fpr$  (Fig.~\ref{fig:fpr-exp}) is an affordable increase before elements are removed, and a potential improvement when element bits are deleted. 
 For other parameters ($m$, $r$, $k$), we could verify the adherence 
 to theory, with the above noted divergences, too. 


\section{Example applications and future work}
\label{sec:applications}

We now give a snapshot on two networking applications to illustrate the practical use of the DlBF when placed
into fixed-length packet headers. 
 In LIPSIN~\cite{lipsin}, the inserted elements are unidirectional link identifiers (LID). 
 A 256-bit source routing BF can be constructed by including the LIDs of a multicast delivery tree. 
 Being able to remove already processed LIDs enables (1) avoiding loops, and 
(2) deleting special LIDs like multi-hop virtual links or control messages. 

In a second scenario, we are exploring the DlBF in a data center environment to compactly represent a sequence of middlebox services (e.g., firewall, load balancer, DPI) which a packet needs to transverse. 
Relying on a substrate of switch programmability (OpenFlow), the content of the DlBF is used to transparently 
forward packets 
upon match on Bloomed Service IDs, which are removed after leaving the middlebox. 

Future work includes exploring \textit{dynamics} along two axes. First, understanding the practical limits if we keep doing insertions and deletions. Second,
investigating a dynamic adaptation of the amount and the bit range of the deletable regions in function on how collisions happen. 
An open question is if there are other compact and more flexible ways to code the information of the collision-free regions.
Finally, the \textit{power of choices} at hashing time may introduce another interesting interplay. 
 For instance, creating \textit{d} DlBF candidates with different sets of hash functions and selecting the best  in terms of $fpr$ or guarantees that certain elements are deletable. 


\section{Conclusion}

This letter introduces the deletable Bloom filter (DlBF), a new Bloom engenderment based on the idea of compactly encoding the information of where collisions happen when inserting elements. This allows safely (i.e. without introducing false negatives) elements removal. Depending on how much memory space one is willing to invest, different rates on element deletability and false positives can be achieved. 
The DlBF is simple and can be easily plugged to existing BFs. 
We briefly presented two packet forwarding applications benefiting from the DlBF, which we believe could be a good fit for existing (and upcoming) friends of the Bloom principle.



\bibliographystyle{IEEEtran}

\bibliography{esteve_CL2010-0344}

\end{document}